# PHENOMENOLOGY OF EARTHQUAKES

A.V. Guglielmi

*Institute of Physics of the Earth, RAS, Moscow*

*guglielmi@mail.ru*

*Abstract*

Phenomenology is the unity of principles and methods of studying the essence of phenomena. This paper is a concise review of recent works in which the phenomenological ideas of physics are used to analyze earthquakes. An example of a phenomenological theory is thermodynamics. Maxwell's electrodynamics is also a perfect example of a phenomenological theory. The phenomenology of earthquakes has not yet reached such a level. So far, we have reached the status of prolegomena for the future phenomenology of earthquakes. The current state can be characterized as preliminary. Nevertheless, in this paper, using specific examples, it is shown that when searching for the foundations of a theory, when processing and analyzing specific manifestations of seismicity, it is useful to use phenomenological concepts of general physics.

*Contents*

## 1. Introduction

This paper is a concise review of the series of works united by the common idea, the essence of which is that phenomenological concepts of physics are used for processing and analysis of earthquakes. A typical task of earthquake phenomenology is to find a mathematical formulation of the hyperbolic law of aftershock evolution, which was discovered by Omori in 1894 [1]. The author drew attention to the fact that the Omori hyperbola is the resolvent of a simple differential equation containing a quadratic nonlinearity [2]. The equation admits interesting generalizations, which make it possible to comprehend in a uniform way a rather wide range of seismic phenomena.

The series of works mentioned above was carried out in recent years by a small team of geophysicists with the participation of the author. The works are partially presented in reviews [3, 4].

There is an opinion that any physical theory is nothing more than phenomenology, since the laws of nature known to us contain more or less arbitrary phenomenological parameters that must either be measured experimentally or calculated on the basis of a more general phenomenological theory. In this sense, the phenomenology of earthquakes is identical to the physics of earthquakes. At the same time, however, it is worth keeping in mind the philosophical aspect of phenomenology, radically expressed by Husserl [5]. In particular, his idea of a phenomenological reduction seems to us useful in the search for a rational theory of earthquakes.

## 2. Omori law and its generalization

We will use the master equation

$$\frac{dn}{dt} + \sigma n^2 = 0 \qquad (1)$$

as the basis for the phenomenological theory of aftershocks. Here $n(t)$ is the frequency of aftershocks averaged over small time intervals, and $\sigma$ is the deactivation coefficient of the earthquake source that "cools down" after the main shock. The reason for our choice of the master equation is the fact that for $\sigma = \text{const}$ the general solution of Eq. (1) coincides with the algebraic form $n = k/(c+t)$ of the classical Omori law [1] if we put $c = \tau_0/\sigma$, $k = 1/\sigma$, $\tau_0 = 1/n(0)$ [2].

For a number of reasons, the differential form (1) is better than the algebraic form of the law proposed by Omori. First, the hard constraint $\sigma = \text{const}$ is removed and the solution of the evolution equation takes the following more general form:

$$n = 1/(\tau_0 + \tau). \qquad (2)$$

Here

$$\tau = \int_0^t \sigma(t')dt'. \qquad (3)$$

Formula (2), like the classical Omori formula, expresses the hyperbolic dependence of the frequency of aftershocks on time, and at the same time, it takes into account that after the occurrence of the main rupture, time in the source, figuratively speaking, flows unevenly. The uneven flow of time is associated with the non-stationarity of the parameters of geological environment in the source, and the non-stationarity can be explained by complex processes of rock relaxation to a new state of equilibrium after the previous state was sharply disturbed by the main shock of the earthquake.

Secondly, equation (1) gives us the opportunity to formulate and solve the inverse problem of the earthquake source. We will consider this aspect in the next section of the paper. Finally, Eq. (1) suggests possible ways of generalizing the law of aftershock evolution. We point out two such possibilities here.

Faraoni [6] drew attention to the fact that equation (1) can be represented as the Lagrange equation and proposed a bold extrapolation of the phenomenological theory of earthquakes. Klain and the author [7] modified the Faraoni Lagrangian and derived the logistic equation

$$\frac{dn}{dt} = n(\gamma - \sigma n) \tag{4}$$

to describe the evolution of aftershocks. Here $\gamma$ is the second phenomenological parameter of our theory. With the help of the logistic equation, a phase portrait of a dynamical system simulating the evolution of aftershocks was constructed, and an attempt was made to comprehend the origin of earthquake swarms and mainshocks.

The following generalization was prompted by an unexpected result of experimental study of the space-time evolution of aftershocks [8]. It was found that aftershock epicenters propagate from the mainshock epicenter with a velocity of several kilometers per second. In order to interpret the propagation effect, the equation of nonlinear diffusion was proposed:

$$\frac{\partial n}{\partial t} = n(\gamma - \sigma n) + D \frac{\partial^2 n}{\partial x^2}, \tag{5}$$

Here the $x$ axis is directed along the earth's surface, and $D$ is the third phenomenological parameter of the theory. At one time, equation (5) was studied by Kolmogorov, Petrovsky, and Piskunov [9]. They showed that the equation has solutions in the form of slow traveling waves (see, for example, [10]).

### 3. Inverse problem

The differential form (1) of Omori's law enables us to formulate and solve the inverse problem of aftershock physics. The inverse problem is to calculate the source deactivation factor from the observational data on the frequency of aftershocks [3, 4, 11–13].

Let us make the change of variable $n \to g = 1/n$ and rewrite (1) in the form $dg/dt = \sigma$.

Formally, we solved the inverse problem, but in practice the solution turns out to be unstable due to strong fluctuations of the original function $n(t)$. Regularization in this case consists in replacing $g \to \langle g \rangle$, where the angle brackets denote the operation of smoothing the auxiliary function. As a result, the solution takes the form

$$\sigma = \frac{d}{dt}\langle g \rangle. \tag{6}$$

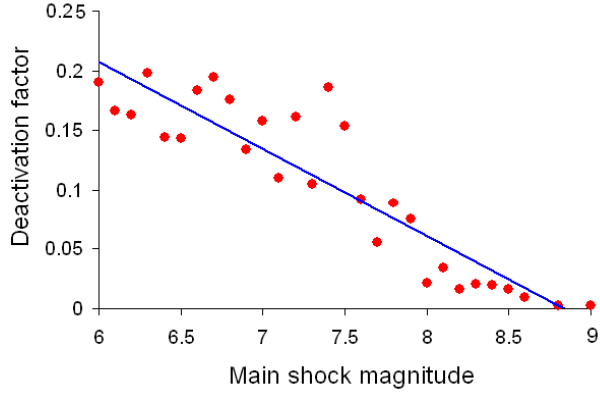

**Fig. 1.** Dependence of the deactivation factor of the earthquake source on the magnitude of the main shock [14].

Experience shows that the deactivation coefficient $\sigma(t)$ undergoes variations over time, but at the first stage of evolution $\sigma = \text{const}$. The corresponding time interval was called the *Omori epoch*. In the Omori epoch, the classical Omori law is fulfilled, according to which the frequency of aftershocks $n(t)$ hyperbolically decreases over universal time *t*. The duration of the Omori epoch varies from case to case from a few days to three months. It was predicted theoretically and then found experimentally that the deactivation coefficient in the Omori epoch is the smaller, the greater the magnitude of the mainshock (figure 1).

## 4. Proper time

Let us return to formulas (2), (3). We call the value of $\tau$ the *proper time* of the earthquake source. The specific dependence of $\tau$ on the universal time $t$ reflects the non-stationarity of the geological environment in the source. Non-stationarity is associated with a complex and difficult-to-control process of relaxation of rocks in the source after the formation of the main rupture.

In reality, $n(t)$ can noticeably deviate from the classical formula $n = k/(c+t)$, while formula (2) for $n(\tau)$, on the one hand, retains the hyperbolic structure of the Omori law, and on the other hand, quite satisfactorily approximates the observational data.

So, in the case of aftershocks, we have a dynamical system (1). The phenomenological parameter $\sigma(t)$ characterizes the source and the outer fields in which it is located. Omori law in the form of differential equation (1) suggested to us how to take into account the latent non-stationarity of a source and external fields using the concept of proper time (3).

Let us show that the idea of proper time can be useful not only in modeling the aftershock flow, but also in studying global seismicity [15, 16]. As an object of study, we will choose the global activity of strong earthquakes ($M \geq 7$). We will present the sequence of strong earthquakes as a Poisson process, i.e. as a chain of instantaneous events separated by random intervals of time. We use comparatively weak earthquakes as the "underground clock", the ticking of which marks the course of proper time. Let's choose earthquakes with magnitudes $6 \leq M < 7$ as a trial run.

The Poisson distribution has the form

$$p_k(\lambda) = \frac{\lambda^k}{k!}\exp(-\lambda). \quad (7)$$

At $\lambda = at$, the $p_k(t)$ value is the probability that $k$ events occur during time $t$, $k = 0,1,2,3,...$. The average number of events is

$$\langle k \rangle = \sum_{k=0}^{\infty} k p_k(t) = at. \quad (8)$$

If we put $\lambda = b\tau$, then we get $\langle k \rangle = b\tau$ in quite the same way. In other words, the Poisson process leads to a linear increase in the accumulated number of events over the course of world and/or proper time.

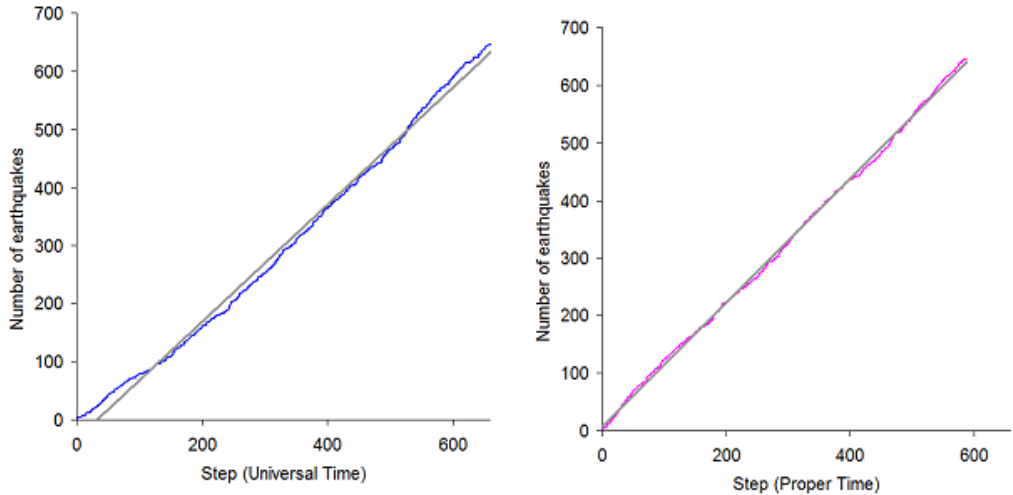

Fig. 2. Dependence of the accumulated number of earthquakes with $M \geq 7$ on world time (left) and proper time (right) [15].

Let's look at figure 2. It shows the dynamics of the accumulation of events with the growth of world time on the left, and with the growth of proper time on the right. The time steps are chosen so that the number of points in the figure on the left approximately coincides with the number of points in the figure on the right. On the left, the step is equal to 26 days, and on the right, 10 strokes of the "underground clock". (See [15] for more details on the construction of the figure.)

We see that, on average, the number of accumulated events increases linearly with the course of both world and proper time. Let us, however, pay attention to the fact that when ordering events according to proper time, the experimental curve deviates less from a straight line than when ordering according to world time.

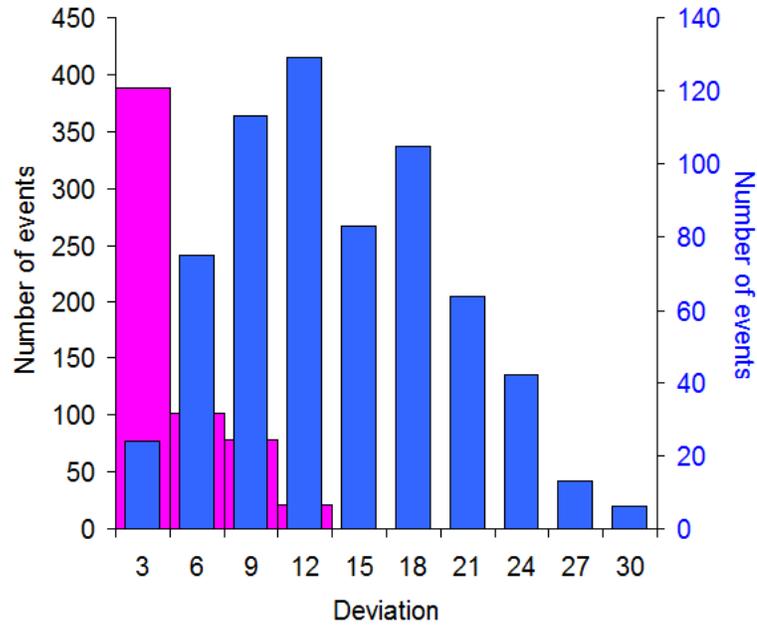

Fig. 3. Distributions of the absolute values of the deviation of the curves shown in figure 2 left (blue) and right (red).

To make this difference more convincing, figure 3 shows the distributions of the deviation modules of real curves from straight lines. It is obvious that using proper time fits better with the theory of the Poisson process than using world time.

## 5. Triggers

It is logical to add the function $f(t)$ to the right side of the master equation:

$$\frac{dn}{dt} + \sigma n^2 = f(t) \tag{9}$$

The inhomogeneous differential equation (9) can be used to model additive triggers that induce aftershocks. Here we will consider two endogenous triggers, one of which is pulsed ($f \propto \delta(t)$), and the second is sinusoidal ($f \propto \sin(\omega t)$) [17–21].

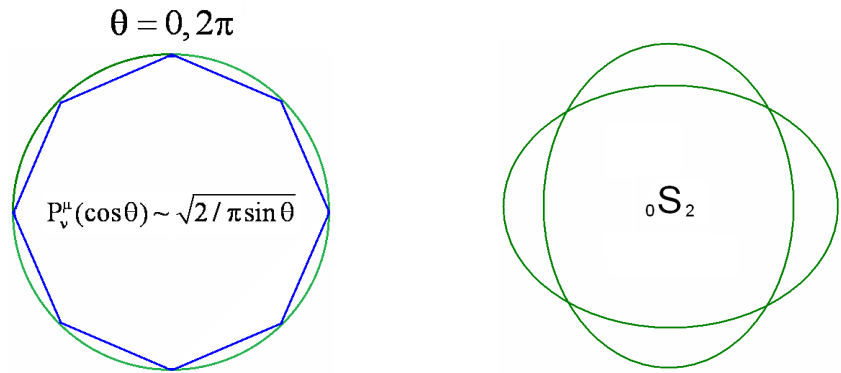

Fig. 4. Schematic pictures of the round-the-world seismic echo (left panel) and the spheroidal oscillations of the Earth (right panel).

The idea is that the main shock excites two nontrivial trigger, namely, the round-the world seismic echo and the free oscillations of the Earth, which may affect the dynamics of the "cooling" earthquake source. Figure 4 illustrates schematically the round-the-world seismic echo and the free spheroidal oscillations of the Earth. The time delay of round-the world echo due to the surface Rayleigh wave is 3 h approximately. The frequency of spheroidal oscillations $_0S_2$ equals 0.309 mHz. The resonant rays of the round-the-world echo formed by the surface and body waves (the smooth and broken lines, respectively). The angular dependence of the short-wave asymptotics of the associated Legendre polynomials, which are proportional to the amplitude of the oscillations, is shown in the central part of the image on the left panel (see [19] for the details).

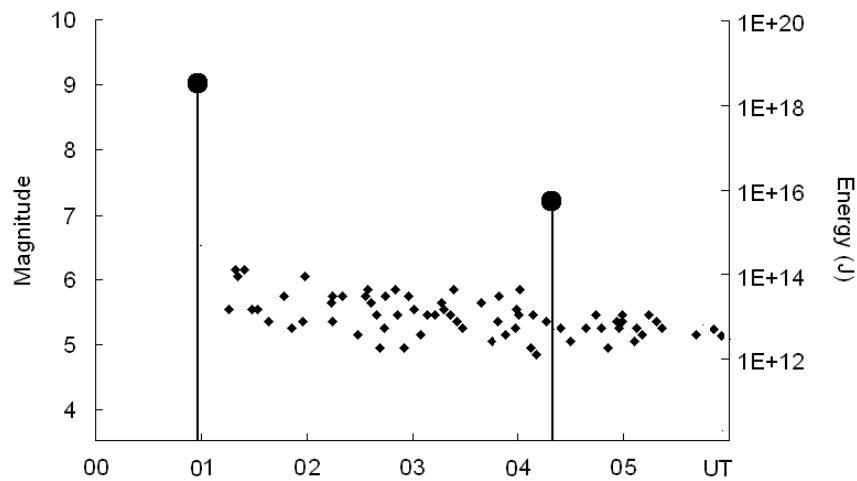

Fig. 5. The Sumatra-Andaman earthquake of December 26, 2004 and its aftershocks within a 5-h interval after the main shock. The black circles mark the main seismic shock and the strongest aftershock [18].

Figure 5 shows the main shock and aftershocks of the Sumatra-Andaman earthquake in the epicentral zone with a radius of 10°. The left vertical axis is scaled in units of magnitude, and the right vertical axis in units of seismic energy. Seventy aftershocks were recorded within a 5-h interval. The strongest aftershock ($M = 7.2$) occurred 3 h 20 min after the main shock. It's clear that this aftershock was induced by the round-the-world seismic echo. Our idea is that the surface elastic waves, which were excited by the main shock, have made a complete revolution around the globe, then returned to the vicinity of the epicenter and induced a strong aftershock there, whose occurrence was energetically prepared by the main shock.

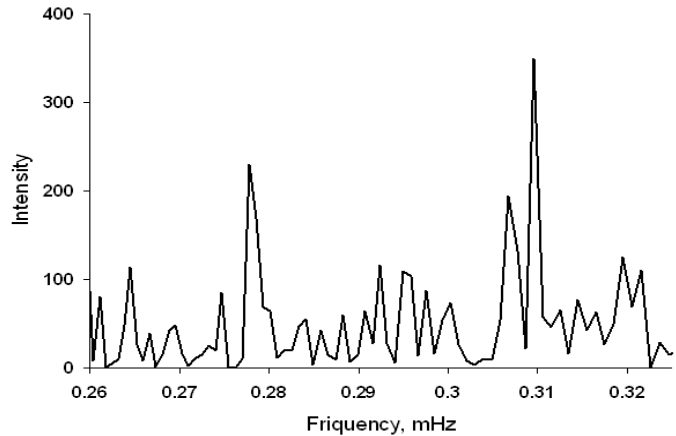

Fig.6. The spectrum of global seismicity (USGS, 1973–2010).

The earthquakes excite the free oscillations of the Earth as a whole at the resonant frequencies of toroidal and spheroidal eigenmodes. Figure 6 shows that global seismicity is modulated by the spheroidal oscillations $_0S_2$, whose frequency is 0.31 mHz.

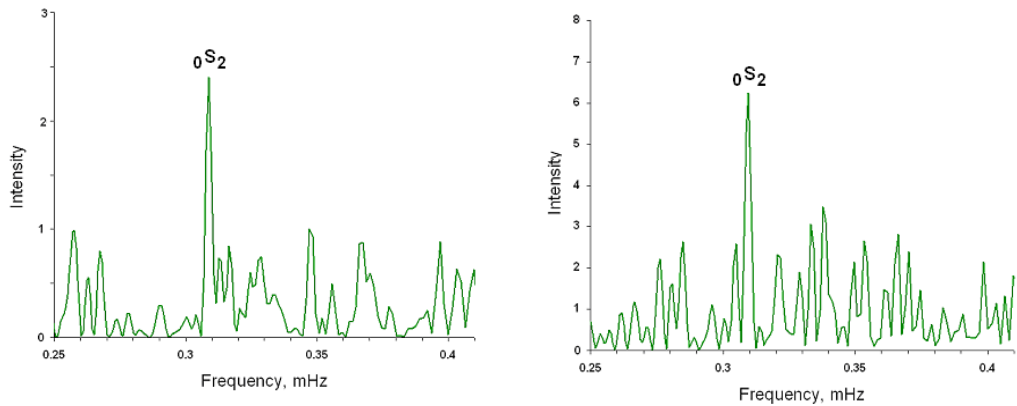

Fig.7. Spectra of aftershocks of two strong earthquakes in California (see the text).

Zotov checked this idea by analyzing the aftershocks in epicentral zones of two strong earthquakes in California. The left panel in Figure 7 shows the spectrum of aftershocks after the mainshock with magnitude $M = 5.9$ occurred 1994.09.12, 12h 23m 43s in Northern California. The right panel shows the spectrum of aftershocks after the mainshock with magnitude $M = 6.0$ occurred 1987.10.01, 14h 42m 20s in Southern California. The aftershocks are analyzed in epicentral zones with radius of 3° during the 80 hours after the mainshocks. We see quite clearly the oscillations of the aftershock activity with the period of $_0S_2$ oscillations of the Earth.

So, earthquakes excite spheroidal oscillations, which, in turn, modulate the activity of earthquakes at the frequency of spheroidal oscillations. The question arises whether the resonant oscillations at the frequency of 0.309 mHz are the self-oscillations of the Earth? From a formal point of view, the answer is yes, since the Earth can be considered an autonomous dynamical system in a good approximation. However, the substantive meaning of the answer is not entirely clear yet.

Recall that the term "self-oscillations" was introduced in 1928, but even Galileo, Hooke and Huygens experimentally studied self-oscillations as a physical phenomenon in connection with the problem of improving clock mechanisms. Rayleigh theoretically described a number of mechanical and acoustic systems capable of generating self-oscillations. He emphasized non-linearity as the most important property of oscillatory systems of this kind. Nowadays, Mandelstam and Andronov have created a school of the theory of non-linear oscillations.

The self-oscillatory system must contain a number of structural elements – a resonator, a source of constant energy, and a positive feedback device. For example, in a mechanical watch, this is a pendulum, a kettlebell or a spring, and the feedback is carried out using a ratchet connected to an energy source and an anchor attached to the pendulum. It is quite obvious that in our case the resonator is the Earth as a whole, capable of performing resonant oscillations at one of its natural frequencies, for example, at a fundamental frequency of 0.309 MHz. The solid shell of the Earth (lithosphere) is constantly in a nonequilibrium stress-strain state. It serves as a source of energy necessary for the excitation of resonant oscillations. It remains to indicate the feedback mechanism, but it is precisely at this point that difficulties and doubts arise, which are analyzed in detail in [22].

At the end of this section, we mention a wide class of anthropogenic triggers. They should be considered exogenous triggers. It is not always clear whether they are additive or multiplicative. The significant impact of anthropogenic triggers on seismicity has been confirmed by numerous observations. The weekend effect [23] deserves special attention, as well as the so-called Big Ben effect [24, 25], which manifested itself in Figure 5 as a peak at the frequency of 0.277 mHz.

## 6. Triads

A sufficiently strong earthquake is called the mainshock if it is followed by aftershocks – earthquakes of lesser strength, and the Bath's law [26] is fulfilled, according to which the maximum magnitude of the aftershocks is less than the magnitude of the mainshock by at least $\Delta M = 1$. Quite often, but not always, the main shock is preceded by foreshocks, the number of which $N_-$ is usually much less than the number of aftershocks $N_+$: $N_- \ll N_+$. A peculiar triunity "foreshocks – mainshock – aftershocks" was proposed to be called the *classical triad* [19]. If there are no foreshocks, then such a "triad" can naturally be called a *shortened classical triad*.

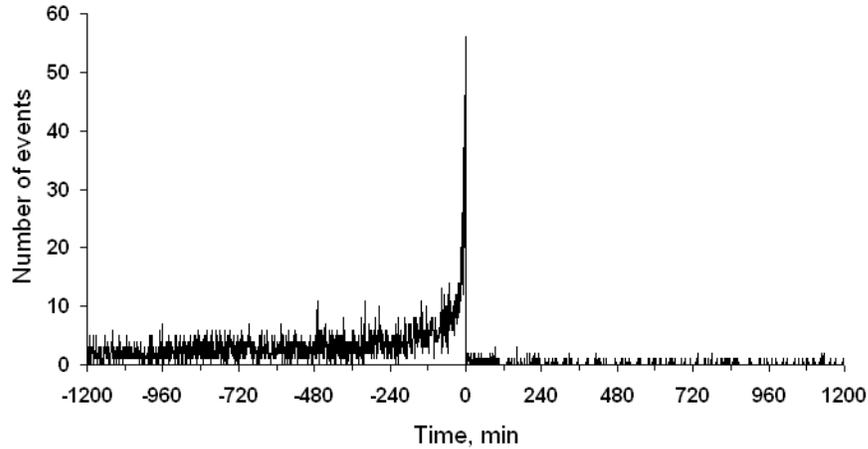

Fig. 8. Time distribution of foreshocks and aftershocks of mirror triads in the range of magnitudes of the main shocks $M_0 = 5-6$.

Zotov posed the question of the existence of mirror triads in which $N_+ \ll N_-$ [27, 28]. He found that the mirror triads form an essential complement to the classical triads. Figure 8 gives an idea of the shape of the mirror triad. Sometimes a shortened mirror triad is observed, in which aftershocks are absent altogether.

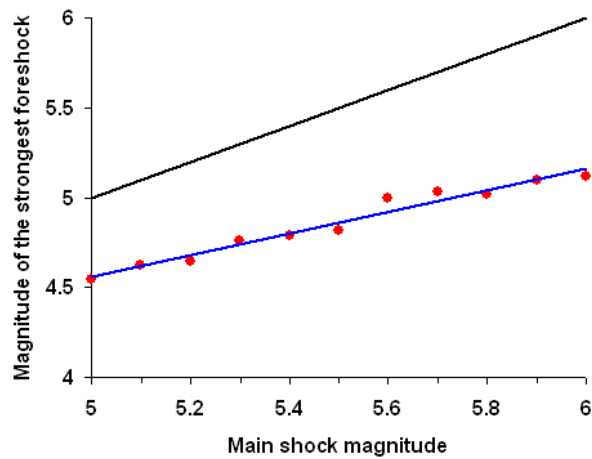

Fig. 9. Comparison of the magnitudes of the strongest foreshocks and the magnitudes of the main shocks [27]. The blue line approximates the experimental points. The black line is drawn for clarity.

This begs the question of whether the strongest earthquake in the mirror triad should be called the main shock? By definition of the main shock in the classical triad, Bath's law is fulfilled. But Bath's law imposes a limit on the magnitude of aftershocks, and there may be no aftershocks at all in the mirror triad. Nevertheless, judging by figure 9, it is possible to give a positive answer to the question posed above. We see that for mirror triads a peculiar analog of Bath's law is valid.

Namely, the minimum gap between the magnitude of the main shock and the magnitude of the strongest foreshock is $\Delta M = 0.5$. (Recall that for aftershocks in the classical triad $\Delta M = 1$.)

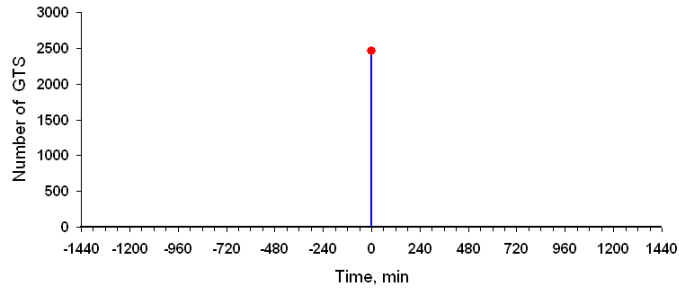

Fig. 10. Solitary earthquakes with magnitudes $M_0 \geq 6$.

Often there are symmetrical triads. In such triads $N_- = N_+$. Finally, a large collection of degenerate symmetric triads was found, in which $N_- = N_+ = 0$ (see figure 10). A fairly strong isolated earthquake of this kind can be called *Grande terremoto solitario*, or GTS for short [4].

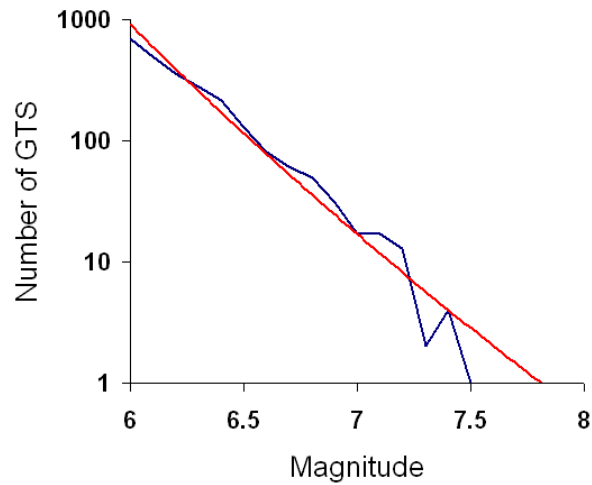

Fig. 11. Dependence of the number of GTS on the magnitude. The Gutenberg-Richter distribution is shown with a red line [27].

The Gutenberg-Richter law [29] seems to hold true for non-classical triads. In particular, the law is uniquely satisfied for the GTS: $\lg N = 10.6 - 1.4 M$ (see figure 11).

So, there are several varieties of main shocks. They differ from each other in the relative abundance of foreshocks and aftershocks. Of particular interest are GTS, which occur spontaneously under very calm seismic conditions, and are not accompanied by aftershocks. This suggests an analogy between the GTS and the so-called "Rogue waves" (or f
"Freak waves") – isolated giant waves that occasionally emerge on a relatively quiet ocean surface.

## 7. Discussion and conclusion

The brief review of the works presented in this paper could be called "Earthquake Geometrodynamics". The term *geometrodynamics* was introduced into theoretical physics by Wheeler to denote the project of constructing a unified field theory in order to consistently describe physical reality in the language of geometry (see, for example, [30, 31]). Geometrodynamics of earthquakes in the form in which it appears to us at the present time does not pretend to solve a problem of this scale. Nevertheless, it is impossible not to notice that when searching for the foundations of the theory of earthquakes, we quite often use geometric images, concepts and representations.

For example, the source boundary can be defined as the minimum convex hull of the hypocenters of all aftershocks. The geometric image of a point naturally arises in the geometrodynamics of the earthquake source when trying to represent something like the "center of gravity" of aftershocks. There is no doubt that as the aftershock excitation process develops, the center of gravity will migrate, reflecting the dynamics of rock relaxation after the formation of the main rupture. The sphericity of the Earth leads to the phenomenon of a round-the-world echo, the cumulative effect of which creates a powerful aftershock three hours after the mainshock. Spheroidal and toroidal oscillations of the Earth modulate global seismic activity. Geometrodynamics also includes the propagation of aftershock epicenters from the epicenter of the mainshock at a velocity of several kilometers per hour, which was found in the experiment. The examples could go on. In other words, in relation to a narrow, but quite definite range of phenomena, the term *geometrodynamics of earthquakes* seems to be appropriate. However, the term used in the title of this paper more fully reflects the specifics of the proposed approach to the study of earthquakes.

We understand phenomenology as the unity of principles and methods for studying the essence of phenomena. By and large, we have been guided, if necessary, by the key provisions of Husserl's phenomenology. The specificity of our approach lies in the methodical use of the phenomenological methods of general physics in the study of earthquakes. Whenever possible, we avoid analogies between an earthquake and the dynamics of a sandpile, a financial disaster, an epidemic, and the like. Analogies of this kind awaken the imagination, give impetus to the emergence of new ideas, but they do not fit into the context of general physics, which has developed effective methods of phenomenological thinking.

An example of a phenomenological theory is thermodynamics. Maxwell's electrodynamics is also a perfect example of a phenomenological theory. Extremely schematizing, we can say that classical electrodynamics consists of two unequal parts – equations describing the evolution of the electromagnetic field, and phenomenological coefficients, such as the dielectric constant. The equations have the status of laws (axioms), while the phenomenological parameters, generally speaking, remain undefined. The fundamental laws of physics, such as the principle of causality and the law of conservation of energy, impose a number of restrictions on the phenomenological parameters, but in general the parameters are more or less arbitrary. The parameters can be measured experimentally, or calculated within the framework of one or another model of the medium.

The physics of earthquakes has not yet reached such a level. The evolution equations do not reflect the whole variety of manifestations of seismicity. And only in rare cases it is possible to measure the phenomenological parameters experimentally. For example, we have learned how to measure the deactivation factor, but we do not have a realistic model in which to calculate, or even estimate in order of magnitude, the deactivation factor. The incompleteness of the phenomenological approach to the study of earthquakes is obvious. The current state of earthquake research can be described as preliminary. So far, we have reached the status of prolegomena for the future phenomenology of earthquakes.

In an expanded form, the paper will be published in the Journal Volcanology and Seismology [32].

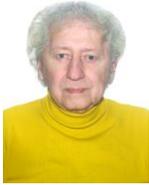

*Author information*. I was born in 1935. I received my education at the Radiophysical Faculty of the University in Gorky (now Nizhny Novgorod). My teachers were Boris Nikolaevich Gershman and Vitaly Lazarevich Ginzburg. Currently, I am a professor, chief researcher at the Institute of Physics of the Earth, RAS. Until 2016, my work was almost exclusively in space electrodynamics. I became interested in the physics of earthquakes under the influence of the research of Oleg Dmitrievich Zotov. Most of the papers in this area I published jointly with him, as well as with Boris Itzikovich Klain and Alexey Dmitrievich Zavyalov. To all of them I express my sincere gratitude.